\documentclass[twocolumn]{revtex4}

\usepackage{graphicx}
\usepackage{bm}
\usepackage{epstopdf}
\usepackage{amsmath,amsfonts,paralist}

\let\a=\alpha \let\b=\beta  \let\g=\gamma  \let\d=\delta \let\e=\varepsilon
     \let\l=\lambda
                
\let\s=\sigma     
   
\let\G=\Gamma   \let\L=\Lambda 
         
  \let\r=\rho

\let\la=\langle
\let\ra=\rangle

\def\ie{{i.e. }}\def\eg{{e.g. }}

\def\CC{{\cal C}} \def\HH{{\cal H}}\def\WW{{\cal W}}
 
  \def\OO{{\cal O}}
 \def\SS{{\cal S}}

\def\to{\rightarrow}
\def\la{\left\langle}
\def\ra{\right\rangle}

\newcommand{\beq}{\begin{equation}}
\newcommand{\eeq}{\end{equation}}

\newcommand{\Tr}{\text{Tr}}

\begin{document}

\title{Critical slowing down exponents in quenched disordered spin models for structural glasses: Random Orthogonal and related models}

%\title{Parisi - Picco - Ritort model at criticality: threshold values of $M$
%and $p$ between continuos and discontinuous transition} 

\author{F.  Caltagirone$^{1,2}$, U. Ferrari$^{1,2}$, L. Leuzzi$^{1,2}$,
G. Parisi$^{1,2,3}$ and T. Rizzo$^{1,2}$} \affiliation{$^1$ Dip. Fisica,
Universit\`a "Sapienza", Piazzale A. Moro 2, I-00185, Rome, Italy \\
$^2$ IPCF-CNR, UOS Rome, Universit\`a "Sapienza", PIazzale A. Moro 2,
I-00185, Rome, Italy \\ $^3$ INFN, Piazzale A. Moro 2, 00185, Rome,
Italy}

\begin{abstract}
An important prediction of Mode-Coupling-Theory (MCT) is the relationship between the power-law decay exponents in the $\beta$  regime.
In the original structural glass context this relationship follows from the MCT equations that are obtained making rather uncontrolled approximations and $\lambda$ has to be treated like a tunable parameter. It is known that a certain class of mean-field spin-glass 
models is exactly described by MCT equations. In this context, the physical meaning of the so called 
parameter exponent $\l$ has recently been unveiled, giving a method to compute it exactly in a static framework.
In this paper we exploit this new technique to compute the critical slowing down exponents in a class 
of mean-field Ising spin-glass models including, as special cases, the Sherrington-Kirkpatrick model, the $p$-spin model and 
the Random Orthogonal model.

\end{abstract}
\date{\today} \maketitle

\section{Introduction and framework}
\label{sec:intro}

It is well known that mean-field spin-glass models have a low temperature phase in which the Replica Symmetry is broken, with 
a breaking pattern that depends on the specific model.
The models displaying a static discontinuous transition which are consistently described by a finite number of breakings are 
characterized by critical slowing down and a dynamical transition at a temperature higher than the static one.

They share some relevant properties of 
structural glasses \cite{Kirkpatrick87a, Kirkpatrick87b, Kirkpatrick87c, Kirkpatrick87d}, more specifically, 
the dynamical equations are exactly equivalent to those predicted by 
the Mode Coupling Theory (MCT) above the mode coupling temperature $T_{mct}$ where ergodicity breaking occurs. \\
The time autocorrelation function in the high temperature phase displays a fast decay to a plateau and then a second relaxation 
to equilibrium. Approaching the dynamical transition temperature (called $T_d$ in the spin-glass context) the length of the plateau 
grows progressively until it diverges exactly at $T_d$, where the system remains stuck forever in one 
of the most excited metastable states in a complex free energy landscape.
According to MCT the approach to the plateau and the decay from it are both characterized by a power-law behaviour, respectively
\beq
\label{eqa}
C(t)\simeq q_d + ct^{-a} 
\eeq
\beq
\label{eqb}
C(t) \simeq q_d - c' t^{b}
\eeq
where $q_d$ is the height of the plateau and the two exponents satisfy the following relation that is exact in the framework 
of MCT (see for example \cite{Gotze})
\beq
\label{exactMCT}
\frac{\Gamma^2(1-a)}{\Gamma(1-2a)}=\frac{\Gamma^2(1+b)}{\Gamma(1+2b)}=\l
\eeq
This relation has been proven to be robust under higher order corrections to standard MCT \cite{Andreanov09}.
The exponent parameter $\l$ and, consequently, the exponents $a$ and $b$ have been computed exactly only for the spherical $p$-spin model \cite{Crisanti93} since the 
dynamical equations are particularly simple and correspond to the so called schematic MCT models.\\
In most of the cases it is, instead, simply considered a tunable parameter, generically connected to the static structure function at $T_{d}$
through an often explicitly unknown functional \cite{Weysser10}.
In the case of continuous transitions, instead, there is no dynamic arrest preceeding the static transition, the time correlation function does not display the two step relaxation and, consequently, no exponent $b$ is defined. At the thermodynamic transition, for long times, the correlation decays to the equilibrium value $q_{EA}$ with a power-law 
of the kind $C(t) \simeq q_{EA} + ct^{-a}$. The equilibrium order parameter $q_{EA}$ is zero at the transition in absence of magnetic field. \\
\indent It has been recently pointed out \cite{Calta11} that there exists a connection between the exponent parameter and the static Gibbs free energy, 
which allows to compute $\l$ in a completely thermodynamic framework, even in cases which go beyond schematic MCT. In the following we will 
briefly summarize the method.

Given a fully-connected model it is possible to compute the Gibbs free energy  $\G(Q)$ as a function of the order parameter that, 
in the case of a spin-glass transition is the well known overlap matrix $Q$. The value of the order parameter can be determined through a saddle point calculation and $\G(Q)$ can then be expanded around this solution. For our ``dynamic'' purposes, the expansion has to be performed around a replica symmetric saddle point solution $Q^{SP}_{ab}=q^{SP}$. This gives raise to eight different kinds of third order terms, but only two of them will be relevant, namely:
\beq
w_1 \Tr (\delta Q^3)= w_1 \sum_{a,b,c} \delta Q_{ab} \delta Q_{bc} \delta Q_{ca}
\eeq
and
\beq
w_2 \sum_{a,b} \delta Q_{ab}^3
\eeq
In the case of discontinuous transitions it can be shown \cite{Calta11} that the following relation holds at the dynamical transition, giving the connection between the dynamical exponents $a$ and $b$ and the static coefficients, namely
\beq
\label{rel}
\l = \frac{w_2(T_d)}{w_1(T_d)}
\eeq 
where $\l$ is given in Eq. (\ref{exactMCT}) the expansion of the Gibbs free energy has to be performed around the value of the overlap yelding the height of the plateau at the dynamical transition, $q_d$. 

Since the coefficients have to be computed at the dynamical transition, where 
quantities at infinite time do {\it not} relax to their equilibrium (thermodynamic) value but remain stuck at their value inside 
the most excited metastable states,
the averages should then be computed {\it inside} a single state. This corresponds to taking a 1-RSB Ansatz with 
breaking parameter $m \rightarrow 1$ or, equivalently, a RS Ansatz 
with the number of replicas $n \rightarrow 1$ \cite{Crisanti08, Monasson95, Franz95a, Rizzo11}. \\
In this paper we will use the second strategy which is technically much simpler than the first one, therefore we cannot treat the case 
of a dynamical transition in presence of a magnetic field, since the mutual overlap ($q_0$) between states is non-zero and 
the 1-RSB ($m\rightarrow 1$)/RS ($n\rightarrow 1$) equivalence does not hold.

On the other hand, for continuous transitions a relation between the exponent $a$ and the two coefficients $w_1$ and $w_2$ analogous to Eq. (\ref{rel}) holds at the static point:
\beq
\l=\frac{w_2(T_s)}{w_1(T_s)}
\eeq
In this case, since the continuous static transition coincides with the dynamical one (\eg in the SK model), the dynamical quantities at infinite time relax to their static value \cite{Sompolinsky82} 
and the averages can be computed in a replica symmetric Ansatz taking finally the limit $n \rightarrow 0$. 
For this reason, if the transition is continuous, the RS Ansatz will be sufficient to treat the case in presence of a magnetic field.

In order to compute the two coefficients $w_1$ and $w_2$ one must determine the expression 
of the Gibbs free energy as a function of the overlap and then expand it to third order around the RS
thermodynamic value $q$.
In fully connected models, introducing a replicated external field $\e$, the free energy reads
\beq
\begin{split}
f(\e)=-\frac{1}{\b n N} \ln \int dQ \exp N\left( \SS[Q] + \Tr \, \e Q\right) 
\end{split}
\eeq
which, for $N \to \infty$, can be evaluated at the saddle point
\beq
\begin{split}
f(\e)=-\frac{1}{\b n } \mathrm{extr}_Q  \left( \SS[Q] + \Tr \, \e Q\right) 
\end{split}
\eeq
We can immediately notice that the equation above exactly defines $f(\e)$ as the {\it Anti Legendre Transform} of the effective action
\beq
\begin{split}
f(\e)=\mathrm{ALT} \, \SS[Q]
\end{split}
\eeq
and, again, by definition the Gibbs free energy $\G(Q)$ is the {\it Legendre Transform} of $f(\e)$, yelding
\beq
\begin{split}
\G(Q)\equiv\mathrm{LT} \, f(\e)= \mathrm{LT} \, \left( \mathrm{ALT} \, \SS[Q]\right) =\SS[Q]
\end{split}
\eeq
This implies that the functional form of the Gibbs free energy is equal to the one of the effective action. In fully connected models, 
we can then directly expand the latter. 
The general form of the third order term in the free energy is
\beq
\label{third}
\SS^{(3)} = \sum_{(ab)(cd)(ef)} W_{ab,cd,ef} \, \d Q_{ab} \d Q_{cd} \d Q_{ef}
\eeq
with
\beq
W_{ab,cd,ef}=\frac{\partial^3 \SS(Q)}{\partial Q_{ab} \partial Q_{cd}\partial Q_{ef}}
\eeq
Since $a\neq b$, $c\neq d$ and $e \neq f$ and the coefficients $W$ are 
computed in RS Ansatz, we can have eight different vertices.
It can be shown \cite{temesvari, Ferrari12} that, restricting the variations to the replicon subspace, \ie the subspace where $\sum_{b} \d Q^R_{ab}=0$, 
one obtains the following expression containing only the two interesting coefficients $w_1$ and $w_2$:
\beq
\begin{split}
&\SS^{(3)}_R = \sum_{(ab)(cd)(ef)} W_{ab,cd,ef} \, \d Q^R_{ab} \d Q^R_{cd} \d Q^R_{ef}=\\
&=w_1\sum_{abc}\d Q^R_{ab}
\d Q^R_{bc}
\d Q^R_{ca}+w_2\sum_{ab}(\d Q^R_{ab})^3
\end{split}
\eeq
that follows quite straightforwardly from equation (\ref{third}) applying the replicon constraint to the variations.\\
In this paper we apply this technique to study the critical slowing down of a general model of mean-field Ising sping-glass 
which includes, as particular cases, the SK model, the $p$-spin model and the Random Orthogonal model (ROM).
The outline of the paper is the following: in section \ref{sec:ising} we introduce the general model, 
in section \ref{sec:computation} we give the details of the computation of the parameter exponent $\l$ for the general case and briefly 
present the result for the SK model and $p$-spin model. In section \ref{sec:rom} we compute $\l$ for the ROM model and compare our exact result with numerical simulations. Finally, in section 
 \ref{sec:summary} we give our conclusions and remarks.

\section{The general model}
\label{sec:ising}
In this section we will consider a class of mean-field models with Hamiltonian
\beq
\label{H_generic}
\HH = -\sum_{i<j} J_{ij} \s_i \s_j - \sum_p \sqrt{\frac{R^{(p)}}{p!} }\sum_{i_1< \dots < i_p} K^p_{i_1, \dots, i_p} \s_{i_1} \cdots \s_{i_p}
\eeq
where $\s_i$ are $N$ Ising spins. The $2$-body interaction matrix is constructed in the following way \cite{cherrier03}:
\beq
J=\OO^{T}\Xi \OO,
\eeq
where $\OO$ is a random $O(N)$ matrix chosen with the Haar measure. On the other hand, $\Xi$ is 
a diagonal matrix with elements independently chosen 
from a distribution $\r(\xi)$. In order to ensure the existence of the thermodynamic limit, 
the support of $\r(\xi)$ must be finite and independent of $N$.
The $p$-body interactions $K^{p}$ are i.i.d. gaussian variables with variance 
\beq
\frac{p!}{N^{p-1}}
\eeq
and 
\beq 
R^{(p)}=\frac{d^p R}{dx^p} (x)\Big{|}_{x=0}
\eeq 
for 
some real valued function $R$.
As shown in \cite{parisiROM, cherrier03, Crisanti92} for this class of mean field spin-glasses, the general form of the replicated free energy is: 
\beq
-n\b f=\mathrm{extr}_{Q,\L}  \SS[Q,\L] 
\eeq
with
\beq
\label{action}
\begin{split}
&\SS[Q,\L] =\frac{1}{2} \Tr \,G(\b Q) + \frac{\beta^2}{2} \sum_{ab} \,R( Q_{ab})  \\
&- \frac{1}{2} \Tr \, Q \L +\ln \left[ \Tr_{\s_a} \, \exp \left( \frac{1}{2} \sum_{a,b} \L_{ab} \,\, \s_a \s_b \right) \right] 
\end{split}
\eeq
where $G:\, M_{n\times n} \to M_{n\times n}$ is a (in general rather complicated) function 
in the space of $n\times n$ matrices, formally defined through its power series around zero. 
The particular form of $G$ depends on the choice of the eigenvalue distribution $\r(\xi)$. In the following 
we will consider mainly two cases: Wigner law and bimodal.\\
Given this effective action, the saddle point equations in $\L$ and $Q$ respectively read
\beq\begin{split} 
Q_{ab} &=\langle \langle \s_a \s_b \rangle \rangle \\
\L_{ab} &= \b [G'(\b Q)]_{ab} + \b^2 R'(Q_{ab}) 
\label{saddle_rom}
\end{split} \eeq
where the average $\langle\langle \cdot \rangle \rangle$ is computed with the measure
\beq
\WW(\L , \s) =\frac{e^{\frac{1}{2} \sum_{a,b} \L_{a,b} \s_a \s_b}}{\Tr_{\s} \,e^{\frac{1}{2} \sum_{a,b} \L_{a,b} \s_a \s_b}}
\eeq
In Replica Symmetric Ansatz ($Q_{ab}=q\,\, , \, \L_{ab}=\hat{\l}$ for $a\neq b$ and $Q_{aa}=q_d\,\, , \, \L_{aa}=\hat{\l}_d$), Eq.s (\ref{saddle_rom}) become
\begin{eqnarray}
\label{saddleRS}
q &=& \langle m^2  \rangle \\
\nonumber
\hat{\l} &=& \frac{\b}{n} [ G'(\b(1+(n-1)q)) -  G'(\b(1-q))] + \b^2 R'(q)
\end{eqnarray} 
where $ m = \tanh (z) $ and the average $\langle \cdot \rangle$ is computed with the measure
\beq
\mu(\hat{\l})=e^{-\frac{z^2}{2 \hat{\l}} } \cosh^n(z) \frac{e^{-n\hat{\l}/2}}{(2\pi \hat{\l})^{1/2}}
\eeq
In the next section we study in detail the (dynamical) critical behaviour of this class of models and we 
show how to compute the critical slowing down exponents.\\

\section{Computation of the MCT exponents}
\label{sec:computation}
As explained in the Introduction, in order to compute the parameter exponent $\l$ we have 
to expand the effective action to third order in $Q$ and then restrict the variations to the replicon subspace, 
obtaining straightforwardly the two coefficients $w_1$ and $w_2$. 

In the present case, the effective action contains the auxiliary field $\L$ which will be eliminated 
making use of the saddle point equation (\ref{saddle_rom}).
The expansion of Eq. (\ref{action}) to third order gives:
\begin{widetext}

\beq
\begin{split}
\d \SS[Q,\L] \simeq &\frac{1}{2} \Tr [ \b G'(\b Q^{SP}) \d Q + \frac{1}{2}\b^2 G^{(2)}(\b Q^{SP}) \d Q \d Q + \frac{1}{3!} \b^3 G^{(3)}(\b Q^{SP}) \d Q \d Q \d Q] + \\
&+ \frac{\b^2}{2} \sum_{ab} [ R'( Q^{SP}_{ab}) \d Q_{ab} + \frac{1}{2} R^{(2)}(Q^{SP}_{ab}) \d Q_{ab}^2 + \frac{1}{3!}  R^{(3)}( Q^{SP}_{ab}) \d Q_{ab}^3]- \\
&- \frac{1}{2} \Tr[Q^{SP} \d \L + \L^{SP} \d Q + \d \L \d Q]+ \frac{1}{2}\sum_{ab}\langle \langle \s^a \s^b \rangle \rangle \d \L_{ab} +\\
&+ \frac{1}{2 \cdot 4}\sum_{ab,cd}\langle \langle \s^a \s^b \s^c \s^d \rangle \rangle_C \d \L_{ab} \d \L_{cd} +\frac{1}{3! \cdot 8} \sum_{ab,cd,ef}\langle \langle \s^a \s^b \s^c \s^d \s^e \s^f\rangle \rangle_C \d \L_{ab} \d \L_{cd} d \L_{ef}
\end{split}
\label{expans}
\eeq

\end{widetext}
A comment is needed for the first line of eq. (\ref{expans}):
the ``scalar like'' Taylor expansion of a matrix functional $f(M)$, $f: \,\,M_{n\times n } \to M_{n\times n }$  around some $M_0$ (different 
from the null matrix), is correct only if $[M_0,\d M]=0$. 

In the present case $Q^{SP}$ is  replica symmetric while $\d Q$ is, in principle, simply symmetric. 
The commutation condition for a RS matrix with a symmetric matrix reads: 
\beq
\sum_{c} \d Q_{cb}=\sum_{c} \d Q_{ac} \,\,\,\, \forall \, a,b
\eeq
that is satisfied in any subspace orthogonal to the anomalous one (see \cite{temesvari}), \ie both in the longitudinal and in the replicon sector \footnote{Any element of the anomalous subspace can be described by a one 
index field, \ie by a vector $\psi^{\a}$ restricted to the condition $\sum_{\a} \psi^{\a}$=0. A generic anomalous field 
can then be written as $\psi^{\a\b}_A = 1/2(\psi^{\a} + \psi^{\b})$. \\
Any element of the longitudinal subspace can be described by a scalar $\psi^{\a \b}=\psi$.}. 
Equating to zero the first order of Eq. (\ref{expans}) we obtain the saddle point equations (\ref{saddleRS}).

Considering that the variations $\d Q$ and $\d \L$ are in the replicon subspace
the second order  
term simplifies as follows
\beq
\begin{split}
& [\b^2 (g^{(2)}_d-g^{(2)}) +\b^2 r^{(2)}]\sum_{ab} \d Q_{ab}^2  \\
& - 2 \sum_{ab} \d \L_{ab} \d Q_{ab} + \langle (1-m^2)^2 \rangle \sum_{ab} \d \L_{ab}^2
\end{split}
\eeq
where, here and in the following formulas we define the four constants (two diagonal and two off-diagonal):
\begin{eqnarray}
\label{equivalence}
\nonumber
g^{(k)}_d &=& \frac{(n-1)G^{(k)} \left( \b (1 - q) \right) + G^{(k)}\left(\b (1 +(n-1) q) \right)}{n} \\
g^{(k)} &=& \frac{G^{(k)}\left(\b (1 +(n-1) q) \right) - G^{(k)}\left( \b (1 - q)\right) }{n} \\
\nonumber
r^{(k)}_d &=& R^{(k)}(1) \\
\nonumber
r^{(k)} &=& R^{(k)}(q)
\end{eqnarray}
and
\beq
\begin{split}
G^{(k)}(x)&=\frac{d^k G(x)}{dx^k}\\
R^{(k)}(x)&=\frac{d^k R(x)}{dx^k}
\end{split}
\eeq
For the system to be critical, the replicon eigenvalue must vanish and, consequently, the Hessian determinant must be zero. 

Imposing this condition we get the following equality
\beq
\label{replicon}
\begin{split}
\langle (1-m^2)^2 \rangle =\frac{1}{[\b^2 (g^{(2)}_d-g^{(2)}) +\b^2 r^{(2)}]}
\end{split}
\eeq
Eq. (\ref{replicon}) together with Eq.s (\ref{saddleRS}) gives the criticality condition, 
which locates the dynamical or static transition point depending on 
the value ($n=1,0$) of the replica number.

Now we want to eliminate the auxiliary field.
The $\L$ - saddle point equation (\ref{saddle_rom}) reads (up to second order)
\beq
\begin{split}
\d Q_{ab} &= \frac{1}{2} \sum_{cd} \langle \langle \s^a \s^b \s^c \s^d \rangle \rangle_C \d \L_{cd} \\
& + \frac{1}{4} \sum_{cd,ef}\langle \langle \s^a \s^b \s^c \s^d \s^e \s^f\rangle \rangle_C \d \L_{cd} d \L_{ef}
\end{split}
\eeq
Exploiting the property of the replicon subspace and the criticality condition we can write the 
variation in the following way:
\beq
\begin{split}
 \d Q_{ab} &= \langle (1-m^2)^2 \rangle \d \L_{ab}  \\
& +\langle (1-m^2)^3 \rangle  \sum_c \d \L_{ac}\d \L_{cb} \\
& + 2 \langle m^2 (1-m^2)^2\rangle \d \L_{ab}^2
\end{split}
\eeq
Inverting the equation we obtain 
\beq
\begin{split}
\d \L_{ab} &=[\b^2 (g^{(2)}_d-g^{(2)}) +\b^2 r^{(2)}] \d Q_{ab} \\
&- [\b^2 (g^{(2)}_d-g^{(2)}) +\b^2 r^{(2)}] ^3  \times \\
&\Bigl(\langle (1-m^2)^3 \rangle \sum_c \d Q_{ac} \d Q_{cb} +\\
&  2 \langle m^2 (1-m^2)^2\rangle \d Q_{ab}^2\Bigr)
\end{split}
\label{lambdasaddle}
\eeq
Now we define
\beq
\begin{split}
&\CC_1\equiv\langle (1-m^2)^3 \rangle \\
&\CC_2\equiv 2 \langle m^2 (1-m^2)^2\rangle
\end{split}
\eeq
and plug the constraint (\ref{lambdasaddle}) into (\ref{expans}), obtaining three different contributions to the third order in $\d Q$, namely
\beq
\begin{split}
& -\frac{1}{2} \Tr[\d \L \d Q] \to \\
&\qquad \qquad \frac{1}{2} [\b^2 (g^{(2)}_d-g^{(2)}) +\b^2 r^{(2)}] ^3 \times \\
&\qquad \qquad \Big(\CC_1 \d Q_{ab} \sum_c \d Q_{bc} \d Q_{ca} + \CC_2 \d Q_{ab}^3\Big) \\
\end{split}
\eeq

\beq
\begin{split}
& \frac{1}{2\cdot 4} (2M_1 -4M_2+2M_3) \sum_{ab} \d \L_{ab}^2 \to \\
&\qquad \qquad -\frac{1}{2} [\b^2 (g^{(2)}_d-g^{(2)}) +\b^2 r^{(2)}] ^3 \times \\
&\qquad \qquad \Big(\CC_1 \d Q_{ab} \sum_c \d Q_{bc} \d Q_{ca} + \CC_2 \d Q_{ab}^3\Big) \\
\end{split}
\eeq

\beq
\begin{split}
&\frac{1}{3! \cdot 8} \sum_{ab,cd,ef}\langle \langle \s^a \s^b \s^c \s^d \s^e \s^f\rangle \rangle_C \d \L_{ab} \d \L_{cd} d \L_{ef} \to \\
&\qquad \qquad \frac{1}{3!}  [\b^2 (g^{(2)}_d-g^{(2)}) +\b^2 r^{(2)}] ^3  \times \\ 
&\qquad \qquad \Big(\CC_1 \sum_{abc}\d Q_{ab} \d Q_{bc} \d Q_{ca} + \CC_2 \sum_{ab}\d Q_{ab}^3 \Big)
\end{split}
\eeq
Summing all the third order contributions in Eq. (\ref{expans}), we eventually find
\beq
\begin{split}
w_{1} &= \frac{1}{2}\frac{\b^3}{3!} (g^{(3)}_d - g^{(3)}) + \frac{1}{3!} [\b^2 (g^{(2)}_d-g^{(2)}) +\b^2 r^{(2)}]^3 \CC_1\\
w_{2} &= \frac{1}{2}\frac{\b^2}{3!} r^{(3)} + \frac{1}{3!} [\b^2 (g^{(2)}_d-g^{(2)}) +\b^2 r^{(2)}]^3 \CC_2
\end{split}
\eeq
Substituting Eq. (\ref{equivalence}) one immediately obtains a general expression for the exponent parameter, that 
is the main result of this paper:
\beq
\label{lambda}
\begin{split}
&\l=\frac{w_2}{w_1}=\frac{ R^{(3)}(q) +  2 \b^4 \mathcal{D}(\b , q)^3 \, \CC_2}{\b G^{(3)}(\b (1-q)) + 2 \b^4  \mathcal{D}(\b , q)^3 \, \CC_1}
\end{split}
\eeq
where
\beq
\mathcal{D}(\b, q)\equiv G^{(2)}(\b (1-q)) +  R^{(2)}(q)
\eeq

\subsection{SK model on the dAT line}
The Sherrington - Kirkpatrick model \cite{SherrKirk75} is described by the Hamiltonian
\beq
\HH = -\frac{1}{2} \sum_{ij} J_{ij} \s_i \s_j - H \sum_i \s_i
\eeq
where the couplings are i.i.d. random variables distributed according to a gaussian with 
zero mean and variance $1/N$.
This model belongs to the class defined above, with $R=x^2/2$ and $G=0$ or, conversely, 
$R=0$ and $G=x^2/2$ \cite{SherrKirk75}, except for the presence of the magnetic field term. 
We will see in a while that this affects the result in a very simple way.\\
It is well known that in the SK model there exists 
a line of instability of the replica symmetric solution in the $\b - H$ plane, 
the de Almeida - Thouless (dAT) line \cite{Almeida}, where the so called replicon eigenvalue of the 
stability matrix vanishes. 
In this section we want to compute 
the decay exponent of the time correlation function along this line.
In order to get the result we first have to find solutions 
which satisfy simultaneously the 
saddle point and the dAT equations, respectively

\beq
q = \int d\mu(z) \tanh^2(\beta \sqrt{q} z + \beta H)
\label{saddleSK}
\eeq

\beq \begin{split}
&1 = \beta^2\int d\mu(z) \,\mathrm{sech}^4(\beta \sqrt{q} z + \beta H) \\	
&d\mu(z)  = \frac{1}{\sqrt{2 \pi}} e ^{- \frac{z^2}{2}}
\label{datline}
\end{split} \eeq

Notice that the expression (\ref{lambda}) is completely general and 
holds for every model belonging to this class, 
while the details of the model enter in the specific form 
of the functions $G$ and $R$. In the case of continuous transitions, the presence of the magnetic field, 
only modifies the definition of the parameter $m$ in equation (\ref{lambda}) which becomes: 
\beq
m=\tanh (z+\beta H)
\eeq
without changing the formal expression for the coefficients. 
Therefore, plugging into the effective action (\ref{action}) $R(x)=\frac{1}{2} x^2$
so that $R^{(2)}(x)=1$ and $R^{(3)}(x)=0$. Then we get immediately 
the expression for the exponent parameter:

\beq
\label{coeffSK}
\l = \frac{\CC_2}{\CC_1}\equiv \frac{2 \la m^2 (1-m^2)^2\ra}{\langle (1-m^2)^3\rangle} 
\eeq

Our result exactly coincides with the one obtained by 
Sompolinsky and Zippelius \cite{Sompolinsky82} in a purely dynamical framework.

\subsection{Multi p-spin Ising model}
Starting from an Hamiltonian of the kind of Eq. (\ref{H_generic}) without
the first term, leads to a generalized version of the
$p$-spin model \cite{Gardner85},
 in wich many multibody interaction terms are considered, depending on
the actual form of the function $R$.
As shown in refs \cite{Crisanti04b,Crisanti07b, Crisanti11},
in these models, the thermodynamic properties, the critical dynamics
and replica symmetry breaking structure depend on the relative
strength of the coupling terms (the coefficients of the expansion of
$R$).
Indeed, in order to treat a particular case, before applying our
technique, one should understand the behavior of the corresponding model.

The simple $p$-spin Ising model is characterized by $R(x)= a x^p$, where a
$a \neq 1$ affects only the variance of the couplings distribution and
indeed rescales the temperature.
For $p>2$, in absence of any external magnetic field, the model displays
a standard RFOT transition from the paramagnetic RS phase to the 1-RSB spin-glass and, at
a lower temperature, a second transition to a FRSB spin-glass \cite{Gardner85}.
Focusing on the first transition, we can compute the critical dynamic
exponents in the present general framework while a specific analysis was presented by some of us in
\cite{Ferrari12}.
In order to recover the exact same model, we have to set $R(x)= x^p/2$,
which reduces Eq.(\ref{saddleRS}) and (\ref{replicon}) to:
\beq
\begin{split}
q &= \langle m^2  \rangle \\
\hat{\l} &=  \frac{p \b^2}{2} q^{p-1}\\
1&=\frac{p (p-1) \b^2 q^{p-2}}{2}\langle (1-m^2)^2 \rangle 
\end{split}
\eeq

and Eq. (\ref{lambda}) to:

\beq \begin{split}
\l=\frac{2 \la m^2 (1-m^2)^2\ra +  \frac{ 2 (p-2) q^{3-2p}}{ \b^4 p^2
(p-1)^2}}{\la (1-m^2)^3\ra}
\label{coeffPspin}
\end{split} \eeq
as it was found in \cite{Ferrari12}.

\section{Random Orthogonal Model}
\label{sec:rom}
\begin{figure}
\includegraphics[width=.99\columnwidth]{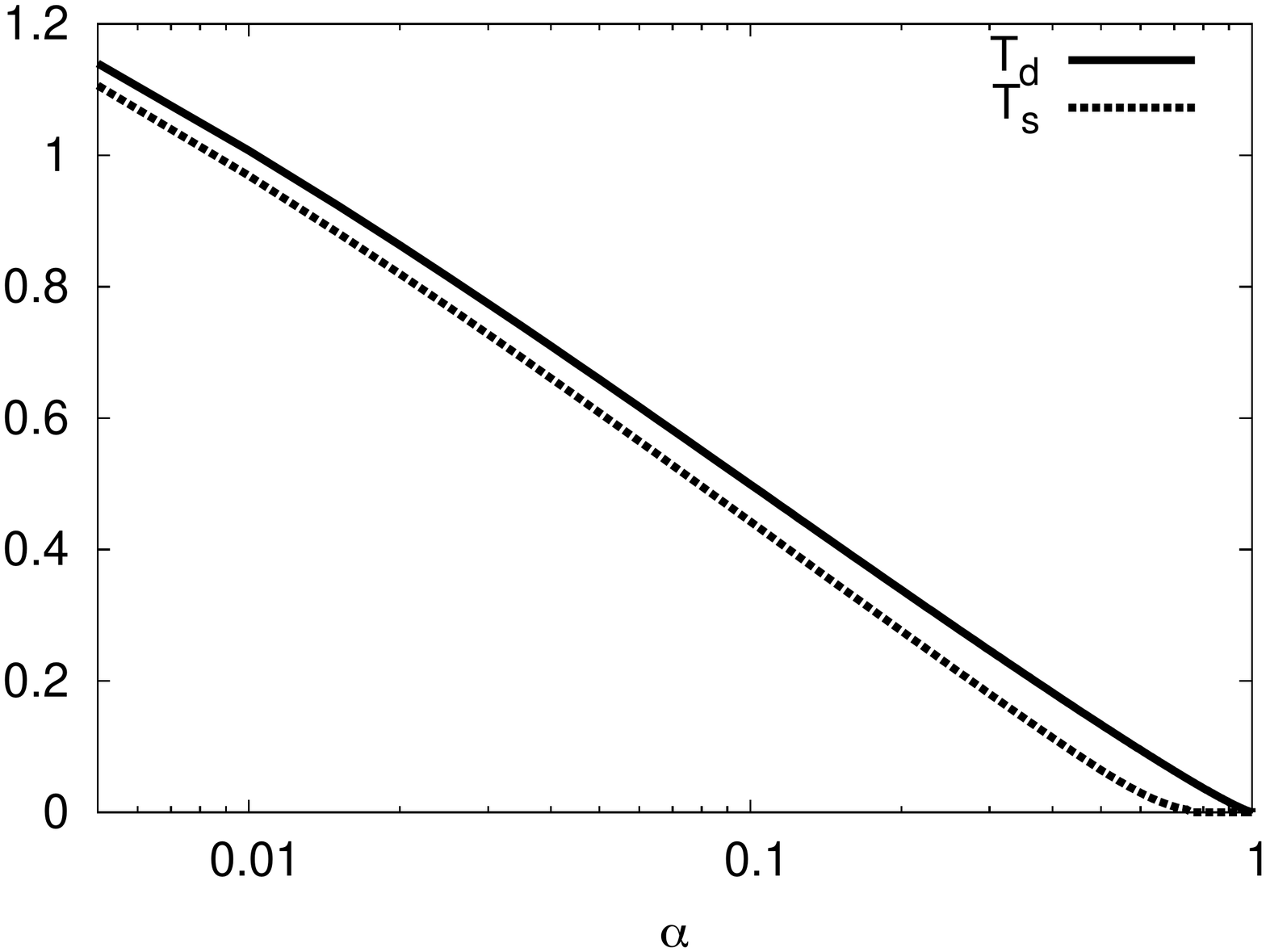}
\caption{The static and dynamical critical temperature as a function of the parameter $\a$.\\
They both diverge for $\a\rightarrow 0$ and are zero at $\a = 1$.}
\label{fig:T_d}
\end{figure}

\begin{figure}
\includegraphics[width=.99\columnwidth]{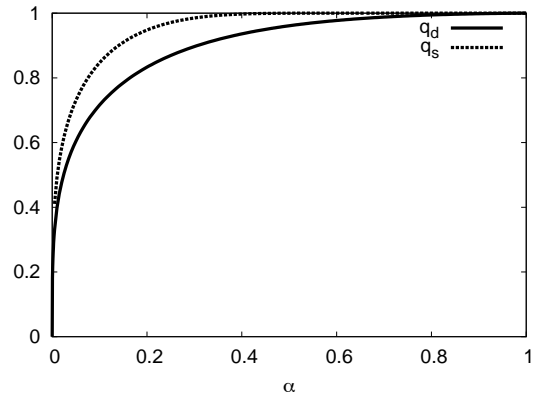}
\caption{The static and dynamical critical overlap as a function of the parameter $\a$.}
\label{fig:q_d}
\end{figure}

The Random Orthogonal Model (ROM) \cite{parisiROM, cherrier03} is obtained with the choice $R=0$ and
\beq
\label{eigen}
\r(\xi)=\a\, \d(\xi - 1) + (1 - \a) \, \d (\xi + 1)
\eeq
It displays a random first order transition regardless of the value of the tunable parameter $\a$. 
The case with $\a=1/2$ has been extensively sudied in \cite{parisiROM} while the general 
case was treated in \cite{cherrier03}.\\
They show that, as a consequence of the choice of the eigenvalue distribution (\ref{eigen}), the function $G$ appearing in the effective action reads \footnote{Note that the corresponding formula in \cite{cherrier03} contains a typing mistake: $\ln(2)$ is 
taken with the negative sign, which would give a negative entropy at infinite temperature.}:
\beq
\begin{split}
2 G(x) &=[1+4x(2\a -1 +x)]^{1/2} \\
&+(2\a -1) \ln \{ [1+4x(2 \a -1+x)]^{1/2} + 2x + 2 \a -1 \} \\
& - \ln \{ [1+4x(2 \a - 1+x)]^{1/2} + 1 + 2(2 \a - 1)x\}  \\
& - (2\a -1) \ln(2\a) - 1 + \ln(2)
\end{split}
\eeq
Using Eq.s (\ref{saddle_rom}) we can determine the transition temperature $T_d$ and the dynamical overlap $q_d$ which 
are shown in Figs. \ref{fig:T_d} and \ref{fig:q_d} and coincide with those found in Ref. \cite{cherrier03}. 

\begin{figure}
\includegraphics[width=.99\columnwidth]{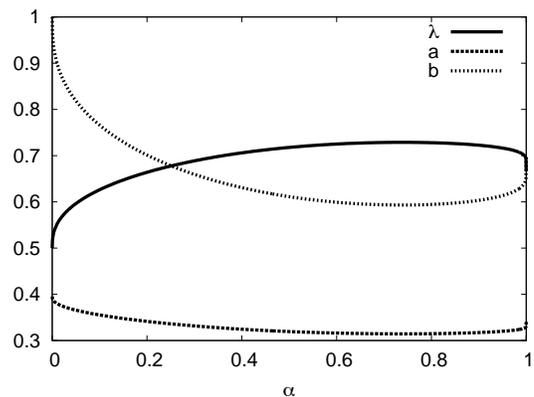}
\caption{Solid line: the exponent parameter $\l$. Dashed line: the exponent $a$. Dot-dashed line: the exponent $b$.}
\label{fig:lambda}
\end{figure}

Once the critical point is obtained as a function of $\a$, using formula (\ref{lambda}) specialized to the ROM case, 
we obtain the value of the exponent parameter $\l(\a)$ and of the critical slowing down exponents $a(\a)$ and $b(\a)$ 
which are shown in Fig. \ref{fig:lambda}. We find numerically that for $\a\rightarrow 1$ the exponent parameter goes to $\l=\frac{2}{3}$ 
as in the Ising $p$-spin model for $p\rightarrow \infty$ \cite{Ferrari12}. 
On the other hand for $\a\rightarrow 0$ we find $\l=\frac{1}{2}$ as in the Ising $p$-spin model for $p\rightarrow 2$.

In particular, for $\a=13/32$ we have $b=0.628$. We now use this value to compare with numerical simulations.

\subsection{Comparison with Monte Carlo data}
\label{sec:results}
There are recent numerical simulations by {\it Sarlat et al.} \cite{sarlat09} on the fully connected ROM which 
give an estimate for the MCT exponents $a$, $b$ and $\g$. 
They choose $\a = \frac{13}{32} \simeq 0.4$ in order to have higher transition temperatures and a good separation 
between the static and dynamical critical temperature. \\
Their direct estimate of the exponent $b$ is $0.62$, while their direct estimate of $\gamma$ is $2.1$ which, through the exact MCT relations: 
\beq
\begin{split}
&\frac{\Gamma^2(1+b)}{\Gamma(1+2b)}=\frac{\Gamma^2(1-a)}{\Gamma(1-2a)} \\
& \gamma=\frac{1}{2a} + \frac{1}{2b}
\end{split}
\eeq
yields $b_{MCT} \simeq 0.75$.\\
Our exact computation yields instead $b_{th}\simeq 0.628$,
which suggests that the best estimate of the exponent $b$ in \cite{sarlat09} is the direct one, that is 
very close to the actual value.

\section{Summary and conclusions}
\label{sec:summary}
In the present work we have introduced a general fully-connected 
model for Ising spins, which combines an orthogonal two body interaction with 
a set of $p$-body interactions. 

Exploiting a technique that has been recently introduced \cite{Calta11}, 
based on the equivalence between statics and long time dynamics,
we have been able to find an analytic expression for the exponent parameter $\l$, 
in a purely static framework. As particular cases of the general model we have studied the Sherrington-Kirkpatrick model 
along the de Almeida-Thouless line, the $p$-spin model and the Random Orthogonal model. \\
\indent For the SK model we find the same result found by Sompolinsky and Zippelius in Ref. \cite{Sompolinsky82}.\\
\indent For the $p$-spin we recover, as a byproduct of the general model, the results given in detail in ref. \cite{Ferrari12}.\\
\indent We have studied the critical behaviour of the parameteric class of Random Orthogonal models at arbitrary values of the 
constant $\a \in [0,1]$, that determines the distribution of the eigenvalues of the interaction matrix. 
The exponent parameter and the two MCT exponents have been determined analitically for any 
$\a$ and in particular we have looked at $\a=13/32$ in order to make a comparison 
with existing numerical simulations \cite{sarlat09}. 
Our exact result is in very good agreement with the one obtained in the Monte Carlo study, through a direct estimate 
of $b$ (late $\b$ regime). \\
\indent On the other hand, a direct estimate of the exponent $\g$ gives a result that is quite far from what we found here, 
suggesting that the strong finite-size corrections affect the value of $\g$ much more than $b$. Numerical interpolations 
at criticality are very sensitive for glassy models and the corresponding estimates can strongly suffer of this drawback.

%%%%%%%%%%%%%%%%%%%%

%\input{Mp_model_mf.bbl}
\bibliography{francescobib.bib}

\end{document}